# Polarization dependent, surface plasmon induced photoconductance in gold nanorod arrays

S. Diefenbach[1], N. Erhard[1], J. Schopka[1], A. Martin[2], C. Karnetzky[1], D. Iacopino[2], and A. W. Holleitner[1]

[1]Walter Schottky Institut and Physik-Department, Technische Universität München, Am Coulombwall 4a, D-85748 Garching (Germany)

[2]Tyndall National Institute, University College Cork, Lee Maltings Complex, Dyke Parade, Cork, Ireland.

E-mail: holleitner@wsi.tum.de and daniela.iacopino@tyndall.ie

ABSTRACT

We report on the photoconductance in two-dimensional arrays of gold nanorods which is strongly enhanced at the frequency of the longitudinal surface plasmon of the nanorods. The arrays are formed by a combination of droplet deposition and stamping of gold nanorod solutions on $SiO_2$ substrates. We find that the plasmon induced photoconductance is sensitive to the linear polarization of the exciting photons. We interpret the occurrence of the photoconductance as a bolometric enhancement of the arrays' conductance upon excitation of the longitudinal surface plasmon resonance of the nanorods.

(PACS: 78.67.Qa, 73.20.Mf, 73.50.Pz)

# 1. Introduction

The excitation of surface plasmons in closely-spaced arrays of metal nanorods leads to a strong enhancement of the electromagnetic fields at the surface of the rods at visible and near-infrared wavelengths [1][2][3][4][5][6]. To date, this phenomenon has been used in three-dimensional disordered arrays for sensing applications [7], due to the enhancement of fluorescence [8][9][10][11][12][13] or Raman signatures of molecules adsorbed in close proximity to nanorod junctions [14][15][16][17][18]. Ordered two-dimensional arrays of metal nanorods have attracted a lot of interest because they are prospective candidates for use in nanophotonic (waveguiding)[19][20][21][22][23][24] and nanoelectronic (molecular junctions) applications [25][26][27].

Generally, there are certain well-known electrodynamic fingerprints of surface plasmons [28]. On the one hand, they can be identified as energy resonances in the optoelectronic excitation spectra of plasmonic structures [29]. On the other hand, surface plasmons have a specific spatial response pattern depending on the geometry of the plasmonic structures. This distribution of the local electromagnetic fields can translate into a polarization dependence of the exciting photon in the far-field. For instance, the surface plasmon mode of spherical nanoparticles splits into a longitudinal and a transversal mode for elongated nanorods [30]. In turn, the excitation of the split modes is strongly anisotropic and polarization dependent [31]. Recently, the photoconductance of two-dimensional arrays of *spherical* gold nanoparticles was explored [32]. The macroscopic dimensions of such well-ordered arrays allow to read-out the excitation of surface plasmon modes in a straight-forward (photo-) conductance measurement. Indeed, the photoconductance of such arrays was found to be enhanced at the photon energy corresponding to the surface plasmon resonance. However, the photoconductance of the arrays with spherical particles did not show any polarization dependence inherently related to the surface plasmon modes [33].

Here, we report on the photoconductance properties of two-dimensional, well-ordered arrays of elongated nanorods made out of gold. They are fabricated by a combination of droplet deposition and stamping methods. We measure the photoconductance of the arrays as a function of voltage, photon energy and the angle of the electric field polarization. We find a clear enhancement of the photoconductance at the photon energy of the longitudinal surface plasmon of the nanorod arrays. In addition, the photoconductance reaches its maximum value when the electric field of the exciting photons is aligned with the longitudinal axis of the nanorods. This polarization dependence proves that the longitudinal plasmon in the nanorods gives rise to an enhanced conductance of the overall array. Microscopically, we interpret the plasmonic photoconductance as a bolometric enhancement of the conductance of the nanorod arrays upon excitation of the surface plasmon resonance. Along this interpretation, we find very good agreement between the data and a simple model which estimates the conductance change of the nanorod arrays due to a plasmonically induced temperature increase. Our results demonstrate that we can obtain local control of the electric transport in a nanorod array by optical means. This can be interesting for applications as optical sensors or for local temperature measurements.

## 2. Materials and Methods

2.1. Materials and Synthesis

Tetrachloroauric acid, cetyltrimethylammoniumbromide (CTAB), sodium borohydride, silver nitrate, ascorbic acid, hydrochloric acid, nitric acid, chlorobenzene, mercaptosuccinic acid, tetraoctylammoniumbromide (TOAB) are purchased from Sigma-Aldrich. All glassware is cleaned with aqua regia prior to the nanorod synthesis. Milli-Q water (resistivity > 18 M$\Omega$ cm$^{-1}$) is used in all experiments. The synthesis of nanorods is performed as follows [34]. For the seed solution, CTAB (5 mL, 0.20 M) is mixed with 5.0 mL of 0.00050 M HAuCl$_4$. While stirring, 0.60 mL of ice-cold 0.010M NaBH$_4$ is added which results in the formation of a pale brown solution. Vigorous stirring of the seed solution is continued for 2 minutes. Afterwards, the solution is kept at 30 °C. For the growth of the nanorods, CTAB (25 mL, 0.20 M) is added to 1 mL of 0.004 M of AgNO$_3$ at 30 °C. To this solution, 25 mL of 0.0010 M HAuCl$_4$ is added, and after gentle mixing of the solution, 0.35 mL of 0.0788 M ascorbic acid is added. Ascorbic acid as a mild reducing agent changes the growth solution from orange to colourless. The final step is the addition of 70 μL of the seed solution at 30 °C. The colour of the solution gradually changes within 10-20 min. The temperature of the growth solution is kept constant at 30 °C until the growth process of the nanorods is completed (90-120 min). In the next step, the Au nanorods in the water solution are centrifuged and re-dispersed in water so that the final CTAB concentration is lower than 0.02 mM. In order to transfer nanorods into chlorobenzene, mercaptosuccinic acid (3 mL, 10 mM) is added to 3 mL of the aqueous nanorod solution. The pH is adjusted to be 9 under vigorous stirring. To this solution, 1.5 mL of TOAB chlorobenzene (50 mM) is added. The resulting mixture is left under vigorous stirring for 30 min until the water phase discolours, and the organic phase becomes intense red.

## 2.2. Device Fabrication

For the fabrication of ordered Au nanorod arrays, a small aliquot (10 µL) of the Au nanorod chlorobenzene solution ([Au] = 10 nM) is deposited on a glass coverslip substrate, covered with a petri dish and left to evaporate at room temperature for 3 h. The controlled evaporation of the solvent results in the formation of domains containing nanorods arranged parallel to the substrate in a side-to-side fashion. The obtained nanorod arrays are transferred onto a host Si substrate with a 150 nm thick $SiO_2$ layer on top. To this end, the original glass substrate is placed onto the second host substrate. Then, the two substrates are pressed together for 20 seconds. This transfer step is required to improve the adhesion of the Au nanorod arrays onto the host substrate. The excess surfactant and organic solvent residues are chemically removed by multiple washes with isopropanol.

In a shadow mask evaporation step, domains of the nanorod arrays are contacted on the host substrate by macroscopic gold electrodes with a separation of ~10 µm. The dimensions of the electronically contacted area of nanorods are approximately 80 × 10 µm$^2$. The gold contacts are electrically connected to a chip carrier with gold wires using a wedge bonder.

## 2.3. SEM Characterisation

Scanning electron microscopy (SEM) images of individual gold nanowires on the first $SiO_2$ substrate are acquired using a field emission SEM (JSM-6700F, JEOL UK Ltd.) operating at a beam voltage of 5 kV (figure 1(a)). Such SEM images of single nanorods allow us to determine the average length and width of the nanorods to be (39.9 ± 3) nm and 11.1 ± 2 nm, respectively. In addition, SEM images of two-dimensional nanorod arrays on the host substrate are taken (NVision40, ZEISS) at a beam voltage of 3 kV. Figure 1(b) shows an SEM image of such a nanorod array, constituted by a solid monolayer phase, ordered both orientationally and positionally. Within this phase, the nanorods are assembled side-to-side in a two-dimensional order.

## 2.4. Optical Characterisation

UV-vis absorption spectrum of gold nanorods dispersed in aqueous solution is acquired using an Agilent 8453E (Agilent Technologies Inc.) spectrophotometer. The optical path length is 10 mm with pure solvent being used as a reference. Maximum absorption corresponding to nanorod longitudinal mode is 754 nm.

## 2.5. Conductance and Photoconductance Measurements

All conductance and photoconductance measurements are carried out at room temperature in a vacuum chamber at a pressure of about $10^{-5}$ mbar. Figure 1(c) shows a typical current-voltage characteristic of the nanorod arrays without laser excitation. The corresponding array exhibits an almost ohmic behavior which reveals a diffuse charge transport regime due to thermal activation [35][36]. For a bias voltage $|V_{sd}| < 2$ V, we determine a two-terminal conductance of 2.9 µS.

The photoconductance $G_{photo}$ of the nanorod arrays is measured as a function of the photon wavelength $\lambda$, a chopper frequency $f_{chop}$, the location of the laser excitation with respect to the source-drain contacts, and the photon polarization. To this end, a bias voltage $V_{SD}$ is applied across the source-drain electrodes, while the light of a Ti:sapphire laser is focused through an objective of a microscope onto the nanorod arrays. For a modulation frequency of ~2 kHz, a reference trigger is given by the chopper frequency $f_{chop}$ of a chopper wheel. The current across the sample is fed through a current-voltage converter. A lock-in amplifier triggered at $f_{chop}$ provides a dc-signal of the current difference $I_{photo} = I_{on} - I_{off}$ across the sample for the laser being "on" or "off". Figure 1(d) depicts a typical dependence of $I_{photo}$ on the applied bias voltage $V_{SD}$. The signal $I_{photo}$ at a certain $V_{SD}$ can be translated into a photoconductance $G_{photo} = I_{photo} / V_{SD}$ of the electronically contacted nanorod arrays because the measured samples show an ohmic behaviour in the examined range of $V_{SD}$. The zero-potential of the current-voltage converter ensures that the source-drain voltage $V_{SD}$ only drops across the sample. The polarization of the photons is controlled by the orientation of a half-wave plate.

## 3. Experimental Results

### 3.1 Wavelength Dependent Photoconductance

Figure 2(a) shows $G_{photo}$ for different excitation wavelengths. We observe a clear maximum at $(786 \pm 35)$ nm (triangle in figure 2(a)). Consistently, the absorbance spectrum of gold nanorods dispersed in a water solution shows an equivalently broadened resonance with a maximum at 754 nm (figure 2(b)) which can be interpreted to stem from the longitudinal surface plasmon mode of the nanorods [30]. Therefore, the strong maximum of the photoconductance suggests that longitudinal surface plasmons play an important role in the generation of the photoconductance in the nanorod arrays. We further find that the photoconductance depends linearly on the laser power (data not shown).

We note that the second maximum of the absorbance at ~ 540 nm (figure 2(b)) represents the transversal plasmon mode of the nanorods. Unfortunately, this transversal mode is not resolvable in our photoconductance measurements because of an overall noise of $\pm 6$ pS in this wavelength range.

3.2 Polarization Dependent Photoconductance

Figure 3 shows $G_{photo}$ as a function of the rotation angle $\phi$ of the electric field $\hat{E}$ of the incoming laser light at the longitudinal surface plasmon resonance ($\lambda_{photon}$ = 780 nm). The long axis of the investigated nanorod array is defined to be the $y$ direction of a reference coordinate system (inset of figure 3(a)). We observe a maximum of $G_{photo}$ for $\hat{E}$ parallel to $y$ and a minimum for $\hat{E}$ perpendicular to $y$ (figure 3(b)). For a nanorod array domain with the nanorods' orientation in $x$ direction (figure 3(c)), $G_{photo}$ has a maximum with $\hat{E}$ parallel to $x$ and a minimum for $\hat{E}$ perpendicular to $x$ (figure 3(d)).

**4. Discussion**

4.1. Optoelectronic Spectrum of the Photoconductance

In order to describe our observations, we start with the Drude-Lorentz-Sommerfeld model. The dielectric function of an isolated Au nanorod is given by $\varepsilon_{NR} = \varepsilon_{1,NR} + i\varepsilon_{2,NR}$, with

$$\varepsilon_{1,NR} = 1 - \frac{\omega_p^2}{\omega^2 + \Gamma^2} + \varepsilon_{1,core} \qquad \varepsilon_{2,NR} = 1 - \frac{\omega_p^2 \Gamma}{\omega(\omega^2 + \Gamma^2)} + \varepsilon_{2,core} \qquad (1)$$

the real and imaginary parts. Here, $\omega_p$ denotes the bulk plasmon frequency of gold and $\varepsilon_{1,core}$ ($\varepsilon_{2,core}$) is the real (imaginary) part of the core electron contribution to the dielectric response of the nanorodσ. $\Gamma$ (P$_{\varepsilon\phi\phi}$) is a phenomenological size-dependent damping constant, which is given by

$$\Gamma(R_{eff}) = \frac{v_F}{l_\infty} + P\frac{v_F}{R_{eff}} \tag{2}$$

with $v_F$ the Fermi velocity in gold, $l_\infty$ the mean free path of conduction electrons in gold, the effective nanorod diameter $R_{eff} = \sqrt{D \cdot L}$ [37], and $P$ a proportionality factor [38]. The absorption of a two-dimensional layer of nanorods in a dielectric medium can be calculated as the absorption $A_{eff}$ of an effective medium and reads as

$$A_{eff} = 1 - \exp(-\kappa d_{NR}) \tag{3}$$

with $d = 11.1$ nm the thickness of a layer of nanorods and the absorption coefficient [39]

$$\kappa = \omega Im[\varepsilon_{eff}]/cn_{eff} \tag{4}$$

The energy of the absorbed photons is $\hbar\omega$, and $c$ is the velocity of light in vacuum, $\varepsilon_{eff}$ and $n_{eff} = \text{Re}\left[\sqrt{\varepsilon_{eff}}\right]$ are the dielectric function and the refractive index of the effective medium. The Maxwell-Garnett effective medium theory predicts [40][41][42]

$$\varepsilon_{eff} = \sum_{j=1}^{3} \frac{(\varepsilon_{NR}(\omega)-1)f\varepsilon_m + (\varepsilon_m - 1)(1-f)[P_j \varepsilon_{NR}(\omega) + (1-P_j)\varepsilon_m]}{f\varepsilon_m + (1-f)[P_j \varepsilon_{NR}(\omega) + (1-P_j)\varepsilon_m]} + 1 \tag{5}$$

with $\varepsilon_m$ the dielectric constant of the medium surrounding the nanorods and $f$ the volume fraction of the gold. The depolarization factors $P_j$ for the three ellipsoid axes $j = A, B, C$ of the rod with $A > B = C$ are defined as

$$P_A = \frac{1-e^2}{2e^2}\left[ ln\left(\frac{1+e}{1-e}\right) - 2e \right] \qquad P_B = P_C = \frac{1-P_A}{2} \tag{6}$$

with the eccentricity $e = \sqrt{1-\left(\frac{A}{B}\right)}$ and the aspect ratio $R = A/B$ [42].

We adjust $f = 0.1$ and $\varepsilon_m = 2.2$ for the nanorod surrounding. We use the proportionality factor $P = 1.4$ [43], $\omega_p = 1.37 \cdot 10^{16}$ Hz$\omega$ [44], $v_f = 1.4 \cdot 10^6$ ms$^{-1}$ [45], $l_\infty = 29.4$ nm [46], and we take the values of $\varepsilon_{1/2,\text{core}}$ according to Ref. [46] Considering the length and width distribution of the nanorods ($39.9 \pm 3$) nm and ($11.1 \pm 2$) nm, see section 2.3), we calculate the maximum of the absorption to occur in the range of ($789 \pm 38$) nm, which is consistent with the photoconductance (figure 2(a)) and absorbance spectrum (figure 2(b)). Hereby, the calculations corroborate the interpretation that the maximum of $G_{\text{photo}}$ at 780 nm results from a longitudinal surface plasmon resonance in the two-dimensional gold nanorod arrays, and that the broadened width of $G_{\text{photo}}$ is caused by an inhomogeneous size distribution of the nanorods.

We point out that it is experimentally challenging to directly measure the absorbance of the two-dimensional nanorod arrays arranged on the Si/SiO$_2$ substrate, because of the dominating contribution of Si. However, this substrate is chosen for the presented experiments, because it provides a supreme signal-to-noise ratio in the photoconductance measurements. Other substrates, such as the SiO$_2$-substrate used to self-assembly the nanorods before stamping (section 2.2.), exhibit a much worse signal-to-noise performance possibly because of optically excited background charges in the SiO$_2$.

Importantly, since we find that the absorbance spectrum of the nanorods in solution (figure 2(b)) agrees consistently with both the photoconductance spectrum of the two-dimensional nanorod arrays fabricated on the Si/SiO$_2$ substrate and the theoretically expected effective absorbance (line in (figure 2(a))), we argue that the above discussion of a dominating longitudinal plasmon is valid.

4.2. Physical origin of the photoconductance

In the following, we demonstrate that the photoconductance maximum at the plasmon resonance is consistent with a bolometrically enhanced conductance. Generally, the increase of the bath temperature $T_{\text{bath}}$ gives rise to an enhanced dark sheet conductance

$$G_{\text{sheet}} = \frac{dI_{\text{sd}}}{dV} \frac{l}{w} \tag{7}$$

of a nanorod array with length $l$ and width $w$. We find an exponential dependence of $G_{sheet}$ on the inverse temperature (data not shown). This is expected for a nanorod array at temperatures above the Coulomb blockade regime with thermally activated diffusive transport [35][36][47]. From a linear approximation of the measurement at 300 K, we deduce the bolometric gradient of the sheet conductance to be $dG_{sheet}/dT_{bath}=(2.11 \pm 0.177)$ nS/K.

The heat generation in metal nanorod arrays is strongly enhanced by excitation of the surface plasmon resonance [48]. In turn, we calculate the temperature increase of the nanorod array due to the absorbed light. We assume that thermal equilibrium is reached on time scales of nanoseconds to microseconds, significantly shorter than the typical illumination times in our experiment ($1/f_{chop} \sim 0.2$ ms).

We equate the absorbed light intensity $I_{abs} = A_{eff} I_{opt}$ with the heat conducted away from the array through the underlying SiO$_2$ layer. Since the heat conductivity of SiO$_2$ is lower by more than a factor 100 compared to the value of Si, we neglect the heat conductance in the bulk of the silicon chip [49]. Furthermore, we do not consider the heat conductance inside the array due to its small cross section. The heat conductance per unit area of SiO$_2$ is given b

$$Q = \lambda \frac{\Delta T}{d_{SiO_2}} \tag{8}$$

with $\lambda$ the thermal conductivity of SiO$_2$, $d_{SiO_2}$ the thickness of the SiO$_2$ layer, and $\Delta T$ the temperature increase in the nanorod array [49]. We evaluate the temperature increase of the nanorod array at the surface plasmon resonance to be $\Delta T = (1.74 \pm 0.177)$ K.

The experimental parameters are $I_{opt} = 2.5 = 2.5\text{e}$ kW/cm$^2$, $A_{eff} = (0.51 \pm 0.08)$ at 780 nm, $\lambda = 1.1$ W/(m·K) for SiO$_2$ [50] and $d_{SiO2} = 150 \pm 10$ nm..

Since the laser spot has a diameter of only 2 μm and the nanorod array has dimensions of 10 μm × 80 μm, we then describe the array by $(5 \pm 0.5)$ times $(40 \pm 1)$ resistors, each of which has an area of $2 \times 2$ μm$^2$. The resistors are connected in rectangular lattice geometry. In the experiment, only one resistor at a time is excited by the laser beam, and in turn, its conductance is increased by $\Delta G = dG_{sheet}/dT_{bath} \times \Delta T = 2.11$ nS/K $\times 1.74$ K $= 3.668$ . We then calculate the conductance of the whole network for one resistor being illuminated to be

$$G_{calc} = \left( \frac{4}{40G_{sheet}} + \frac{1}{40G_{sheet} + \Delta G} \right)^{-1} - G_{dark} \qquad (9)$$

We determine a photoconductance value at 780 nm of $G_{calc} = (147 \pm 37.2)$ pS. The latter value is in very good agreement with the experimentally determined value of $G_{photo} = 140$ pA/ 1 V $=140$ pS at 780 nm and $I_{opt} = (2.5 \pm 0.05)$ kW/cm$^2$ (figure 1(d)). For the experimental conditions of figure 2(a), we find $G_{calc} = (265 \pm 67)$ pS, which is again in very good agreement with the measured $G_{photo} = (279 \pm 49)$ pS at 780 nm.

Based on this calculation, the photoconductance maximum at the plasmon resonance is consistent with a bolometrically enhanced conductance of the nanorods. We note, however, that similar results on *spherical* nanoparticles were interpreted in terms of a mere plasmonic field enhancement within the nanoparticle arrays [51][52], in terms of trap state filling dynamics in the nanoparticle cores after a plasmonic [53] and a molecular [54] excitation or in terms of a possible hot-electron emission induced by plasmons [55]. In principle, our results at a typical acquisition time scale of $\sim 1/f_{chop} \sim 0.2$ ms do not exclude the possible occurrence of an additional much faster emission of hot electrons which is enhanced through a plasmonic excitation. However, to detect such ultrafast dynamics in the photoconductance more advanced time-resolved optoelectronic circuits need to be applied [56].

Most importantly, the polarization dependence of the $G_{photo}$ in figures 3(b) and 3(d) nicely confirms the interpretation that the photoconductance is induced by the longitudinal plasmon in the nanorod arrays. In the picture of a bolometrically induced photoconductance, the dielectric function (equation (1)) and the absorption coefficient (equation (3)) give rise to an anisotropic, polarization dependent absorption process and therefore heat generation in the nanoparticle arrays. To observe the presented polarization dependence, the laser spot needs to be smaller than the domain size. In the investigated nanorod arrays, the domain size of oriented nanorods is on the order of 1 to 5 □m diameter. By the polarization dependence of $G_{photo}$ on the nanorods' orientation, we can further exclude a possible slit-antenna effect induced by the electromagnetic impact of the two electrodes separated by ~10 μm [33].

## 5. Conclusion

In summary, we report on the photoconductance properties of two-dimensional arrays of gold nanorods which are formed by a combination of droplet deposition and stamping methods. The photoconductance of the nanorod arrays is strongly enhanced through the excitation of longitudinal surface plasmons. Hereby, the photoconductance is found to be polarization dependent with a maximum signal when the electric field of the exciting photon is aligned with the longitudinal axis of the nanorods. We interpret the observations by a bolometric enhancement of the conductance of the nanorod array.

**Acknowledgements**

We gratefully acknowledge support from the European (FP7) 263091 project "HYbrid molecule nanocrystal assemblies for photonic and electronic SENSing applications" (HYSENS).

**Figure 1.** (a) Scanning electron microscope (SEM) image of discrete gold nanorods deposited on $SiO_2$ substrate (scale bar: 200 nm). (b) SEM image of a nanorod array obtained by droplet deposition/stamping on $Si/SiO_2$ substrate (scale bar: 200 nm). (c) Current-voltage characteristic of a nanorod array without laser excitation at room temperature. (d) Photon induced current $I_{photo}$ as a function of source-drain voltage $V_{sd}$ ($f_{chop}$ = 1716 Hz, $I_{opt}$ = 2.5 kW/cm$^2$, $\lambda_{photon}$ = 780 nm ).

**Figure 2.** (a) Open squares: Photoconductance spectrum ( $f_{chop}$ = 3978 Hz, $I_{opt}$ = 4.5 kW/cm$^2$, $V_{sd}$ = 1.0 V). Line: calculated effective absorption $A_{eff}$ according to equations (1)-(6) for $\varepsilon_m$ =2.2, $P$ = 1.4, $f$ = 0.1, $R_{eff}$ = 21.04 nm). (b) UV-Vis spectrum of gold nanorods dispersed in aqueous solution.

**Figure 3.** (a) SEM image of an array of nanorods with orientation along *y* (scale bar: 200 nm). Inset: Orientation of the electric field $\hat{E}$ of the laser. (b) $G_{photo}$ as a function of the rotation angle $\phi$ of $\hat{E}$ for array in (a) ( $f_{chop}$ =5951 Hz, $I_{opt}$ = 2.5 kW/cm$^2$, $V_{sd}$ = 2.0 V, $\lambda_{photon}$ =780 nm) . (c) SEM image of a nanorod array with orientation along *x* (scale bar: 200 nm). (d) $G_{photo}$ vs. $\phi$ for array in (c) ($f_{chop}$ =5951 Hz, $I_{opt}$ = 2.5 kW/cm$^2$, $V_{sd}$ = 2.0 V, $\lambda_{photon}$ = 780 nm).

**Figure 1**

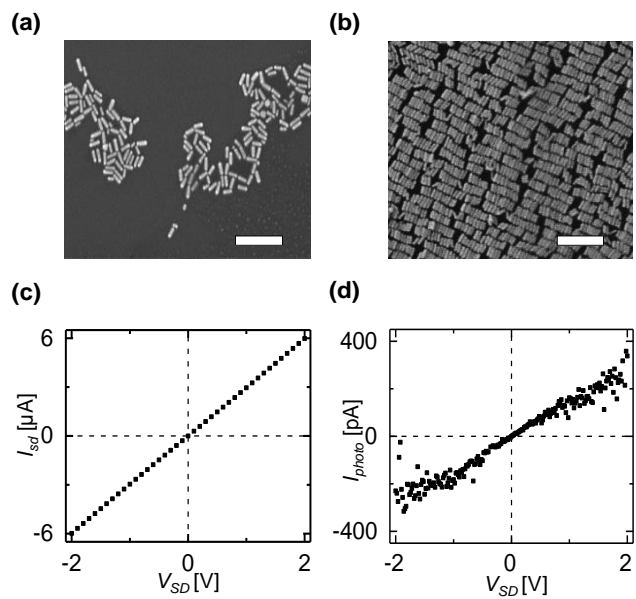

**Figure 2**

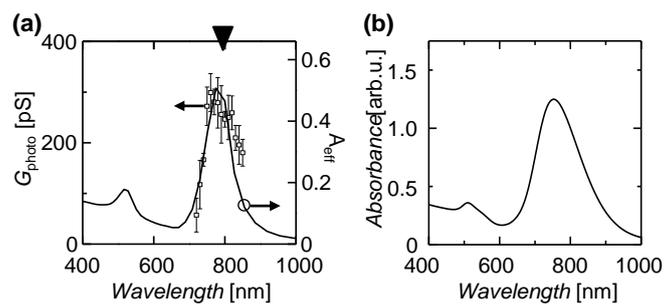

**Figure 3**

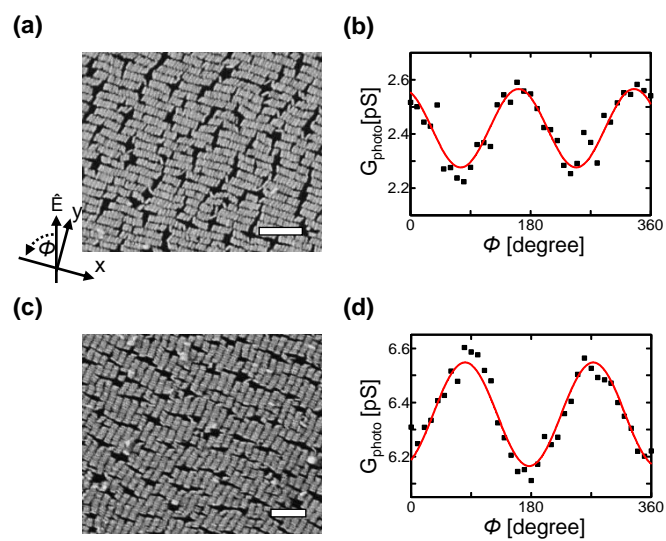